# Stochastic chaos: An analog of quantum chaos

Mark M. Millonas

Complex Systems Group, Theoretical Division and Center for Nonlinear Studies,
MS B258 Los Alamos National Laboratory, Los Alamos, NM 87545
& Santa Fe Institute, Santa Fe, NM

**Abstract**

Some intriging connections between the properties of nonlinear noise driven systems and the nonlinear dynamics of a particular set of Hamilton's equation are discussed. A large class of Fokker-Planck Equations, like the Schrödinger equation, can exhibit a transition in their spectral statistics as a coupling parameter is varied. This transition is connected to the transition to non-integrability in the Hamilton's equations.

In this paper we will be concerned with diffusion processes on $\Re^n$ described by the set of coupled stochastic differential equations

$$dq^i(t) = -\partial^i \Phi(\mathbf{q})dt + \sqrt{g}dW^i(t), \quad i = 1,...,n, \quad (1)$$

where $\Phi(\mathbf{q})$ is a potential bounded from below, the $W^i(t)$ are uncorrelated Wiener processes, and $g$ is a diffusion coefficient. In this case the evolution of the probability density $\rho(\mathbf{q},t)$ on $\Re^n$, is described by the Fokker-Planck equation

$$\partial_t \rho = \frac{g}{2}\Delta \rho + \boldsymbol{\nabla} \cdot (\rho \boldsymbol{\nabla} \Phi). \quad (2)$$

Using the time separation ansatz $\rho(\mathbf{q},t) = \rho(\mathbf{q})e^{-\lambda t/g}$, we can write Eq. (2) as an eigenvalue equation $\mathcal{L}\rho_\lambda(\mathbf{q}) = -\lambda \rho_\lambda(\mathbf{q})$, where $\mathcal{L} = \frac{g^2}{2}\Delta + g\nabla^2\Phi + g\boldsymbol{\nabla}\Phi \cdot \boldsymbol{\nabla}$. After the change of basis $\rho(\mathbf{q}) = e^{-\Phi/g}\Psi(\mathbf{q})$ we obtain

$$\mathcal{H}\Psi_\lambda(\mathbf{q}) = \lambda \Psi_\lambda(\mathbf{q}), \quad (3)$$

where $\mathcal{H} = -e^{\Phi/g}\,\mathcal{L}\,e^{-\Phi/g} = -\frac{g^2}{2}\Delta + \hat{\Phi}(\mathbf{q})$, is a Hermitian Schrödinger type operator with the transformed potential $\hat{\Phi} = \frac{1}{2}(\boldsymbol{\nabla}\Phi)^2 - \frac{g}{2}\boldsymbol{\nabla}^2\Phi$. The problem of solving Eq. (2) has been reduced to the problem defined by equation (3).

For small $g$ the WKB solutions of equation (3) are given by

$$\Psi_\lambda(\mathbf{q}) = \sum_\alpha c_\alpha |\boldsymbol{\nabla} S_\alpha|^{-\frac{1}{2}} \exp\left(\frac{i}{g} S_\alpha(\mathbf{q}, \lambda)\right), \quad (4)$$

where the $S_\alpha(\mathbf{q}, \lambda)$ are the solutions of the Hamilton-Jacobi equation $\frac{1}{2}(\nabla S_\alpha)^2 + \hat{\Phi} = \lambda$. The solutions of this equation are given by the integrals $S_\alpha(\mathbf{q}, \lambda) = \int^\mathbf{q} \mathbf{p}_\alpha \cdot d\mathbf{q}$ where the integration in along the classical trajectories of Hamilton's equations of motion

$$\dot{\mathbf{p}} = -\frac{\partial H}{\partial \mathbf{q}}, \quad \dot{\mathbf{q}} = \frac{\partial H}{\partial \mathbf{p}}, \quad (5)$$

with

$$H(\mathbf{p}, \mathbf{q}) = \frac{1}{2}\mathbf{p}^2 + \hat{\Phi}(\mathbf{q}). \quad (6)$$

The time reversal symmetry of equations (5) with Hamiltonian (6) insures that the eigenfunction (4) are real since the solutions of the Hamilton-Jacobi equation will come in pairs $\pm S_\alpha$. The dynamics of (5) determine the solution of (3) through (4), and the solutions of (2) are given by

$$\rho_\lambda(\mathbf{q}, t) = \exp\left(-\frac{\Phi + \lambda t}{g}\right) \Psi_\lambda(\mathbf{q}). \quad (7)$$

Thus the properties of the Fokker-Planck equation (2) are connected to the dynamics of the system with Hamiltonian (6) in a manner somewhat analogous to the relation of a quantum mechanical system to its classical counterpart.

One question we might ask is how the behavior of (2) is affected by the degree of chaos in the equations of motion (6). Such effects, in the quantum mechanical case ((5) affecting (3)), are often referred to as *quantum chaos*, which is usually defined as the characteristics of quantum systems whose classical analogues exhibit chaos. The statistical properties of the eigenvalues of such systems are such characteristics, and the level spacing distribution $P(S)$, giving the probability of level separation $S$ (measured in units of the local mean spacing), provides one such statistical property. Berry & Tabor[1] have shown that nearly all quantum systems whose classical analogues are integrable will have a Poisson level spacing distribution $P(S) = \exp(-S)$, indicating the statistical independence of neighboring energy levels. On the other hand, it is now understood that the eigenvalues of systems whose classical analogues are chaotic exhibit level repulsion. That is, $P(S) \to 0$ as $S \to 0$.[2] It is expected that systems with time-reversal symmetry whose classical analogues are globally chaotic will have a Wigner level spacing distribution, $P(S) = \pi S/2 \exp(-\pi S^2/4)$, indicating a linear level repulsion as $S \to 0$.[3] Since the eigenvalues of the Fokker-Planck operator, $\mathcal{L}$, with potential $\Phi$ are the negative of the eigenvalues of a Hamiltonian, $\mathcal{H}$, with potential $\hat{\Phi}$, the spectral statistics of the Fokker-Planck equation (2) would then be expected to provide a signature of the dynamics of the equations of motion (5). To explore these ideas Millonas and Reichl[4] studied a family of two-dimensional Fokker-Planck equations with potentials

$$\Phi_\epsilon(x,y) = 2x^4 + \frac{3}{5}y^4 + \epsilon xy(x-y)^2. \tag{8}$$

The system needed to be at least two-dimensional in order to observe chaos in equations (5). When $\epsilon = 0$ the system is completely integrable, since it decouples into two one-dimensional systems. They observed the transition (as $\epsilon$ is varied) in the level spacing statistics of the Fokker-Planck operator as the dynamics of equations (5) changes from completely integrable ($\epsilon = 0$) to almost globally chaotic ($\epsilon = 0.14$). *Stochastic chaos* can then be defined, at least for the case of diffusion in a time-independent potential, as *the properties of stochastic systems described by Eq. (2) when the equations of motion (5) exhibit chaos.* In particular, given a family of potentials $\hat{\Phi}_\epsilon$ where the dynamics of (5) varies from globally integrable to globally chaotic as $\epsilon$ is increased, we would expect the spectral spacing distribution of the $\lambda$'s to exhibit a corresponding transition from Poisson to Wigner level spacing statistics.

An entirely separate problem is the question of the direct physical relevance of the dynamics of (5) to the underlying microscopic dynamics as described by (1). One thing is clear: chaos in (5) is emphatically *not* related to chaos in the dynamics generated by (1) with $g = 0$, that is $\dot{\mathbf{q}} = -\boldsymbol{\nabla}\Phi(\mathbf{q})$. When there is no noise the individual trajectories just follow the gradient of the potential along the route of steepest descent stopping at any local minimum in $\Phi$, so what would normally be considered the underlying microscopic dynamics is trivial, and never chaotic. Thus, there is no simple physical relationship between the dynamics of (5) and the dynamics of (1). A deeper analysis shows that eqs. (5) are the imaginary-time equations of motion for the most probable, or optimal trajectories. These ideas can be extended to the case where there is no detailed balance, but in that case no meaningful analytic continuation of the most probable trajectories is possible. There are still however the optimal trajectories which obey a set of Hamilton's equations with Hamiltonian $\mathcal{H} = 1/2(\mathbf{p}-\mathbf{A})^2 + 1/2\nabla\cdot\mathbf{A}$, where $\mathbf{A}$ is the nongradient force field which replaces $-\partial^i\Phi(\mathbf{q})$ in Eq. 1 in the case where there is no detailed balance. The long-time behavior of systems with or without detailed balance can be calculated in the low-noise asymptotic limit from a knowledge of the optimal trajectories with energy $\lambda = 0$. These are the instanton trajectories which line on the unstable manifold of the Hamiltonian system. This manifold is smooth as a consequence of the center manifold theorem, so chaos will not play a role. However this manifold may have singular projections onto the configuration space in the case where detailed balance is broken, resulting in a rich nonlinear behavior of nonequilibrium stationary states.[5] The most surprising result presented here is than even in the case where there is detailed balance, and the underlying dynamics is completly integrable, chaos will play a role in determining the time-dependent properties of such systems. It appears that there is a deep analogy between quantum dynamics and stochastic dynamics through their relationship to the properties of these conservative dynamical systems. This connection, once made, opens up the study of stochastic processes to a whole range of new tools and concepts.